\documentclass[12pt,fleqn]{article}
\usepackage{palatino,graphicx,wrapfig}
\usepackage[small,compact]{titlesec}
\usepackage{amstext,amsmath,amsfonts,amssymb,color}
\usepackage{epstopdf}
\usepackage{todonotes}
\usepackage{tabularx}
\usepackage{bm}
\usepackage{morefloats}
\usepackage[inline]{enumitem}
\usepackage{soul} 

\usepackage[labelfont=bf,font=normalsize]{caption}
\usepackage[square,numbers,sort&compress]{natbib}

\usepackage{float}
\usepackage{longtable}

\newcounter{arabic}

\newcommand{\tu}{\widetilde{U}}
\newcommand{\tz}{\widetilde{Z}}
\newcommand{\tphi}{\widetilde{\phi}}
\newcommand{\orho}{\overline{\rho}}
\newcommand{\bfphi}{\text{\boldmath $\phi$}}
\newcommand{\bfpsi}{\text{\boldmath $\psi$}}
\newcommand{\bfv}{\bf v}
\newcommand{\mbf}{\boldmath}

\newcommand{\paral}[1]{\paragraph{#1}}

\newcommand{\cored}[1]{{\color{red} #1}}

\newcommand{\squishlist}{
 \begin{list}{$\blacklozenge$}
    { \setlength{\itemsep}{0pt}      \setlength{\parsep}{3pt}
      \setlength{\topsep}{3pt}       \setlength{\partopsep}{0pt}
      \setlength{\leftmargin}{1.5em} \setlength{\labelwidth}{1em}
      \setlength{\labelsep}{0.5em} } }

\newcommand{\squishlisttwo}{
   \begin{list}{$\Rightarrow$}
    { \setlength{\itemsep}{0pt}    \setlength{\parsep}{0pt}
      \setlength{\topsep}{0pt}     \setlength{\partopsep}{0pt}
      \setlength{\leftmargin}{2em} \setlength{\labelwidth}{1.5em}
      \setlength{\labelsep}{0.5em} } }

\newcommand{\squishend}{
    \end{list}  }

\usepackage{dcolumn}
\usepackage{authblk}
\usepackage{subfigure}
\usepackage{subfigmat}
\usepackage{fancyvrb}
\usepackage[lflt]{floatflt}

\begin{document}
\begin{center}
{\bf \Large Classification and Simulation of Anomalous Events in Turbulent Combustion \vspace{0.2in}\\}
{\large Malik Hassanaly, Stephen Voelkel and Venkat Raman \vspace{0.1in} \\}
{\large Department of Aerospace Engineering, University of Michigan \\}
\end{center}
\begin{abstract}

In practical combustion systems, the high nonlinearity of the turbulent combustion process can result in unexpected behavior even for nominal operating conditions. This includes flame flashback in premixed gas turbines, engine unstart in scramjets, and failure of engines to ignite at high altitudes. Many of these events have catastrophic consequences. Hence, understanding the nature of such irregular excursions of combustion systems is important to ensure robust operation. The focus of this work is to identify the sources of such rare excursions from a dynamical systems perspective. Specifically, such anomalous behavior can be explained in terms of three different classes based on the trajectories that the system traverses in its phase-space. For each class, computational algorithms and validation exercises are briefly discussed. 
\end{abstract}

\section{Introduction: The issue of rare events in turbulent combustion}
\label{sec:intro}

The design of practical combustion devices is often an optimization issue, pitting reliability and safety against efficiency and emissions. Depending on the end application, each factor manifests in the design in a different manner. In general, understanding the stability limits and transient behavior of the combustor is important. For instance, high-altitude relight is an important certification test for aircraft engines. Here, cold fuel and air compressed to lower pressure are injected into the combustor, and ignited using an ignition source. The design should ensure that ignition happens within a required time, and that irrespective of the operating conditions, ignition is feasible. Consequently, flame kernel development under cold conditions is a key physical issue in the design of plasma-based ignitors. In land-based gas turbines that predominantly operate in the premixed combustion mode, the issue of thermoacoustic instabilities is of central concern \cite{poinsot_accoustic,candel_accoustic}. Here, the combustor should not amplify and sustain perturbations to inflow and boundary conditions, for a range of load and operating conditions. In both cases, the robustness of the system to variations in inflow or operating conditions needs to be understood. Hence, identifying the propensity of a system to reach unwanted conditions is of critical importance in design. 

In the above context, the focus of this work is on so-called rare events, which are defined as excursions of the system to regions of unlikely and sometimes catastrophic outcomes. A more precise definition is provided later on. Rare events in these systems can for now be thought of as unlikely behavior for a given set of operating conditions. For instance, if the time-to-ignition for a combustor is determined to be $\tau$ for a given set of conditions ${\bm \phi}$, but the system exhibits much shorter or longer ignition times in a particular operation, then such a behavior is termed as a rare event. Such rare events are often associated with catastrophic consequences, for instance, the loss of propulsive power in this case. The rarity of this rare event is, of course, subjective and can be quantified based on a probability density function, which is obtained by posing the ignition time as a probabilistic parameter. Essentially, given the same operating conditions ${\bm \phi}$, the probability of ignition time greater than $\tau_i$ is given by $P(\tau > \tau_i)$. The definition of the rare event can then be based on a threshold probability (say, for instance, $10^{-6}$). 

The fundamental question then relates to the source of such rare events. More precisely, why do systems exhibit such variability even for nominally identical operating conditions? The answer lies in the highly nonlinear dynamics that drive turbulent combustion. It is well-established that turbulent processes are highly chaotic, with small changes to flow conditions resulting in large changes in flow state after a particular time. However, if this is true for all combustion systems, they still operate reliably in vast majority of situations. Consequently, it is important to understand the fundamental nature of such events, and the systems or conditions under which they can be found.

In the broader field of science, there has been a renewed focus on the study of such rare events, based on theories originating in financial markets \cite{blackSwan_taleb, sornette_dk,sornette_dk2}. In particular, the notion of Black Swans \cite{blackSwan_taleb} and Dragon Kings \cite{sornette_dk} are gaining traction. This work extends such theories to the field of turbulent combustion, and intends to build the necessary basis of the rare event analysis from a physical rather than empirical or statistical context. In particular, the rare events are described using a deterministic mathematical formulation well suited for most physical problems.

For turbulent combustion system, it is our hypothesis that such rare events are tied to existence of fluid memory \cite{k65}. By this, it is meant that perturbations made to the system persist for time-scales comparable to large-scale flow time-scales. Of fundamental importance to memory is the existence of low-velocity and low-Reynolds number regions. Although the study of turbulent combustion has focused mostly on the high Reynolds number limit, the stability of most modern combustors depends on low-velocity zones within these devices. For instance, large scale recirculation zones are responsible for flame stabilization in gas turbine combustors. Scramjets operating in the low supersonic regime (Ma $<$ 7) use cavity-based flameholders to stabilize reactions under supersonic or highly compressible flow regimes. In higher flight velocity regimes, recirculation zones behind the injectors are used to stabilize the mixing process. In all of these cases, the loss of a robust flame inside the combustor is related to the destruction of the low-velocity zone. In the limit of high Reynolds numbers, perturbations to the system at any scale will be dissipated very quickly \cite{k59}. In other words, if small perturbations are introduced, they tend to be distributed to scales close to the perturbation scale, and the cascade either above or below this scale is exponentially inhibited. However, as the Reynolds number decreases, these perturbations persist for longer times, and propagate over larger length scales.

The focus of this work is to identify the different modes by which a functional combustion system can be pushed towards an unlikely and therefore unwanted state. The rest of the paper is organized as follows. In the first section, the dynamical systems perspective of the governing equations are introduced. Next we provide a more detailed definition of a rare event.
The following section introduces the classification of anomalous events for this dynamical system. Finally, the classification is used to describe the computational and experimental requirements for developing models that can predict these anomalous events. 

\section{Dynamical system formulation}
\label{sec:dynamical}

A dynamical systems perspective is used for the description of rare events. This approach is particularly useful to understand the continuous drift of a solution towards an anomalous event.

\subsection{Governing equations}
\label{sec:governingEq}

The governing equations for fluid flow and species transport can be written as 
\begin{equation}
    \frac{\partial \rho}{\partial t} + \nabla \cdot \rho {\bm u} = 0,
\end{equation}
\begin{equation}
    \frac{\partial \rho {\bm u}}{\partial t} + \nabla \cdot \rho {\bm u} {\bm u} = -\nabla P + \nabla \cdot {\bm \tau},
\end{equation}
and
\begin{equation}
    \frac{\partial \rho {\bm \phi}}{\partial t} + \nabla \cdot \rho {\bm u} {\bm \phi} = \nabla \cdot \rho D_{\phi} \nabla {\bm \phi} + {\bm S}({\bm \phi}),
\end{equation}
where ${\bm u}$ and ${\bm \phi}$ represent the velocity and scalar fields. It is assumed that the flow configuration can be described using $N_s$ species mass fractions, and an energy equation that is assumed to take the same form as the species transport equation (for the sake of simplicity, since it does not limit the discussion below). $P$ represents pressure, $\rho$ is the fluid density, ${\bm \tau}$ is the Newtonian stress term, $D_{\phi}$ is the vector of species diffusivities, and ${\bm S}$ is the chemical source term. In this study, it is assumed that these equations are solved directly without any turbulence models, implying direct numerical simulations.

The above continuous partial differential equations (PDEs) are discretized in space using appropriate finite-difference schemes, leading to the following set of ordinary differential equations (ODE):
\begin{equation}
\frac{d \bm \xi}{dt} = {\cal N}({\bm \xi},\mathcal{P});~~{\bm \xi}(t=t_0) = {\bm \xi}^0,
\label{eq:dynm}
\end{equation}

where ${\bm \xi} =\{\xi_1, \xi_2,\cdots \xi_M\}$ denotes the vector of variables that define the evolution of the continuous PDEs under consideration, and $\mathcal{P}$ defines the set of parameters necessary to close the system of ODEs. Typically, it denotes the effective macroscopic operating conditions of a gas turbine like the mean inlet and outlet temperature, the bulk flow velocity and the pressure ratio, and the boundary conditions of the system. In using these ODEs, the infinite-dimensional system defined by the PDEs has been restricted to a finite-dimensional state-space. This is justified based on the observation that turbulent structures exhibit a range of length-scales tied to non-dimensional quantities such as Reynolds and Damk\"{o}hler numbers, and viscous dissipation will restrict the smallest scales possible for a given flow. 

\begin{figure}[ht]
\centering
\includegraphics[width=0.55\textwidth]{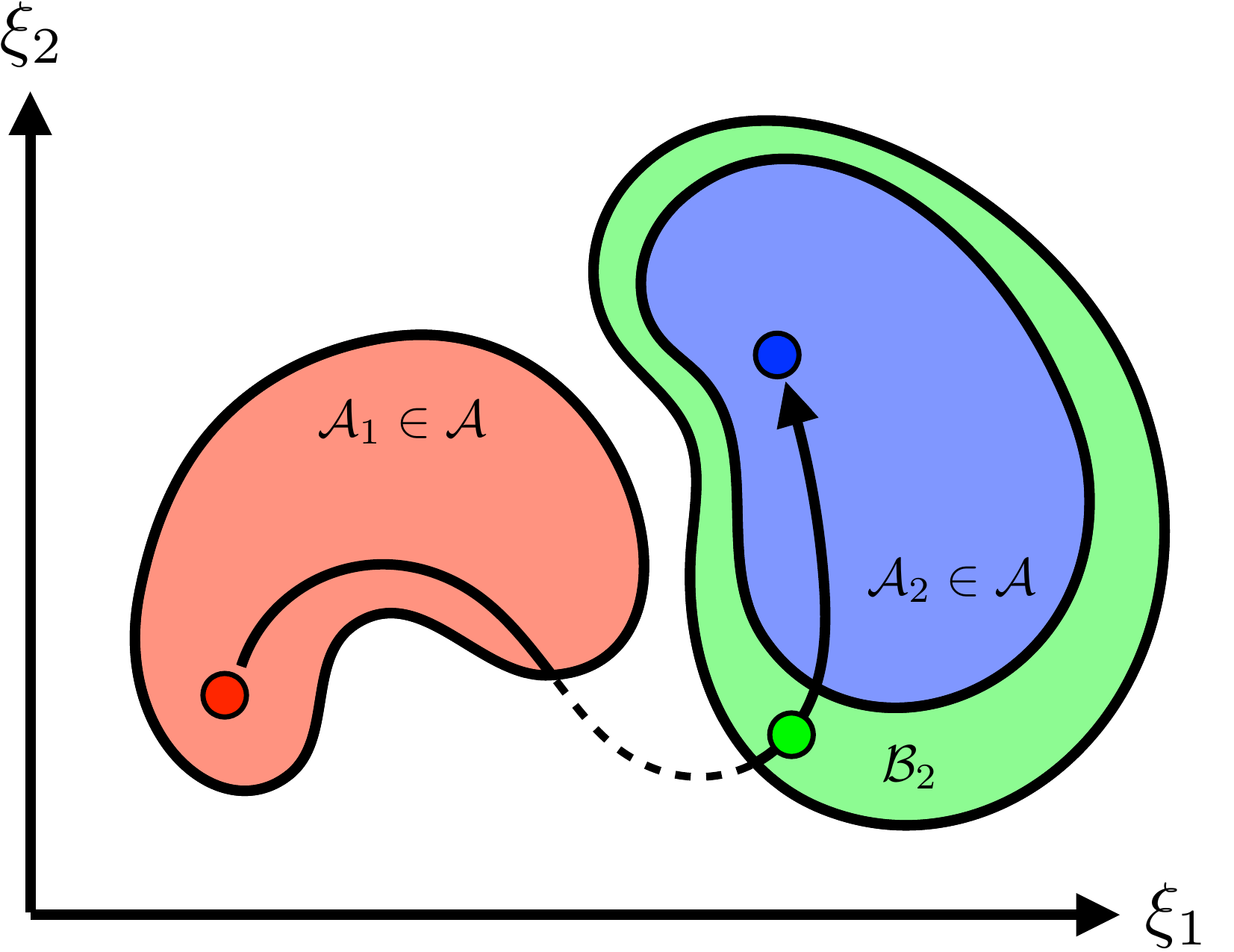}
\caption{Illustration of disjoints attractors. The attracting set $\mathcal{A}$ is the union of two attractors $\mathcal{A}_1 \cup \mathcal{A}_2$. Illustration of Type IIIB event (See Classification section) as an orbit (black line) starting from the red dot, being displaced (dashed line) to the green dot located in the basin of attraction of $\mathcal{A}_2$ noted $\mathcal{B}_2$. The trajectory is eventually attracted to $\mathcal{A}_2$.}
\label{fig:type3b}
\end{figure}

Given this system, a state-space (also called phase-space) can be defined, with dimension $M$ (equal to number of variables), which describes all realizable combinations of variables. The evolution of the combustion system can then be considered as a trajectory moving through this phase-space starting from a given initial condition. The long terms dynamics of the system are described by the ergodic theory and involve the concept of attracting regions of the phase-space. Although multiple definitions for such attracting regions exist \cite{milnor,ruelle81,bonifant}, the vocabulary introduced by Ruelle \cite{ruelle81} is used here. After sufficient time, the solution orbit can lie in the system's attracting set. The attracting set is allowed to be made of disjoint irreducible sets (attractors). For turbulent combustion systems, this means that there can exist hard limits between each part of the attracting set. The limit can be understood as a thermodynamic constraint which forbids to travel from one part of the attracting set to another. However, the solution can lie in each one of the irreducible sets. This property of attracting sets is illustrated in Fig.~\ref{fig:type3b}. The attracting set is defined as the union of two disjoint attractors.

This attracting set itself will span a sub-space of the realizable region, spanning $d$ dimensions. The shape of the attractor is not known {\it a priori}, and cannot be easily extracted using physical observations due to its high dimensionality. Algebraic approximation techniques have been devised \cite{foias_attractor} and used for low dimensional cases only \cite{malykh}. An estimate on the size $M$ can be obtained using Kolmogorov's theory, and scales as $\text{Re}^{9/4}$ \cite{temam}. For reacting flow systems, additional scales due to thin reaction fronts or pressure waves could introduce other scaling relations as well. However, the actual dimension of the attractor is expected to be much smaller due to the strong thermodynamic constraints acting on the system. For example, in most combustion systems, there will be strong correlation between species mass fractions, which is the basis for many combustion models \cite{maas,pitsch_arfm}. Hence $d <M$ for such problems. Regardless, $d$ is sufficiently large that a discrete or analytical characterization of the attracting set is intractable.

\section{Cause of Anomalous Events}
\label{sec:def}
Prior to classifying the anomalous events, it is important to understand the relation to rare events.  Rare events are events intuitively defined as unlikely, which means that the impact of the rare event is not evaluated yet. On the other hand, an extreme event is related to a large departure of a quantity of interest (QoI) from the mean behavior. The extreme event is intrinsically measured according to some metric. The extreme event does not need to be rare but simply remarkable in its variation from the expected QoI. Finally, if the QoI is selected as a projection of the phase-space into a low-dimensional sub-space, multiple events in phase-space can lead to extreme events in the QoI space.

Given the deterministic description of the system provided in the previous section, it is legitimate to ask why the prediction of an anomalous event is an issue. In other terms, if the event is not random, why is it uncertain? As mentioned in the introduction, the answer lies in the variability in the system which can be introduced through three different sources:
\begin{itemize}
    \item A perturbation in the initial or boundary conditions. For instance,
    \[
     {\bm \psi} = \{{\bm u, \bm \phi}\};~~ {\bm \psi}({\bf x}, t=0) = {\bm \psi}^0 + {\bm \epsilon},
     \]
     where ${\bm \epsilon}$ is a perturbation. It is noted that this perturbation need not be small for the discussion below.
    \item A perturbation in the effective operating conditions or a change in the physical domain ($\mathcal{P}$). In other words, not only can the values that the variables take at the boundaries change, but the physical location of the boundary itself might be affected, for instance, due to damages incurred in operation. Similarly, the operating conditions may be perturbed as well.
    \item A perturbation in the fields that momentarily invalidated the equations ($\mathcal{N}$) describing the system. This event would never occur with a complete description of the physical system. This point is only relevant if the solution is driven in a particular direction while violating the governing equations. This can be the result of the addition of perturbations not represented by the governing equations.
\end{itemize}

\section{Classification of Anomalous Events in Turbulent Combustion Systems}
\label{sec:class}

The classification system developed below is geared towards understanding the modeling and simulation requirements for different types of anomalous behaviors in complex reacting flow systems. This classification uses a given set of observable and controllable macroscopic conditions to determine the nature of rare or anomalous behavior in systems. For this purpose, consider ${\cal I} = \{{\cal I}_1,{\cal I}_2,\cdots,{\cal I}_s\}$, the set of controllable input conditions with $s$ distinct components. In general, $s \ll M$, and ${\cal I}$ will be a set of macroscopic input or operating conditions such as pressure, mass flow rate, or fuel-split in multi-injection combustors. The set ${\cal I}$ is precisely known since it constitutes the theoretical set of operating conditions (i.e. the set of conditions the operator wants to enforce). This set is different from ${\cal P}$ which is the set of variables closing the governing equations. Further, consider ${\cal O} = \{{\cal O}_1, {\cal O}_2,\cdots,{\cal O}_q\}$, the set of quantities of interest (QoI) that is used to assess the state of the device. Note that $q \ll M$ also, and will consist of sensor measurements. For instance, this set could include temperature at combustor exit, wall pressure at select locations in the flow path etc. When an anomalous event is detected, additional actuation mechanisms can be built in to return the device to stable state. For instance, flow-bleeding can be used to control unstart in scramjets. 

\paral{Type I events: Predictable Anomalous events -} The Type I events represent reliable behavior of complex devices. Here, for any given ${\cal I}$, the output ${\cal O}$ is precisely known. In other words, there is a direct connection between the input state and the output state. When such a behavior is observed, it is possible to develop a map that relates input to output variables, either through experimental or computational procedures. Once this map is known, the device can be operated within boundaries such that unwanted output states are not observed. For instance, when thermoacoustic instabilities are noted for particular operating conditions, the device can be operated away from these conditions. For most combustion systems, such an operating map is devised in order to understand the limits of stability.

In the sense of the above definition, Type I events do not create rare behavior, but the operating conditions $\mathcal{I}$ themselves could lead to unwanted regimes, and where potentially catastrophic behavior is possible. Most studies of combustion instabilities, and transient phenomena in combustion devices, have focused on Type I events due to their easy reproducibility. Type I events can be readily studied using experiments, since the outcomes are directly dictated by the operating or boundary conditions at a macroscopic scale (such as pressure, inflow velocity, boundary layer thickness etc.), and can be precisely controlled and/or measured.

As discussed before, the dimension of ${\cal I}$ is much smaller than that of the dynamical system. The fact that these inputs are sufficient to guarantee the output state shows that either \begin{enumerate*} [label=\itshape\alph*\upshape)]
\item there occurs a drastic reduction is the true dimensionality of the system, or \item  the output variables ${\cal O}$ are insensitive to much of the state-space of the dynamical system
\end{enumerate*}.
Due to this direct relation between input and output, any variability in the specification of ${\cal I}$ is included within Type I definition.

\paral{Type II events: Uncertainty-driven Rare events -} Type II events are related to the imprecise control over operating and boundary conditions, and may form a superset of Type I events. Type II events assumes a constant value for $\mathcal{P}$ and come from the fact that the set ${\cal I}$ does not fully determine the initial condition ${\bm \xi}^0$ which in turn affects the predictability of ${\cal O}$. In this case, the trajectory, ${\bm \xi}(t)$, is not precisely known, and hence there is ambiguity in the quantities of interest, ${\cal O}$. It is then possible to define a measure in phase-space, which determines the probability of encountering a particular ${\bm \xi}$. Defined as probability $P({\bm \xi})$, the following relation holds:
\begin{equation}
P({\bm \xi}) = P({\bm \xi} | {{\bm \xi}^0}) P({{\bm \xi}^0}),
\end{equation}
where the probabilities on the right hand side are the conditional probabilities of encountering a particular phase-space location given a set of initial conditions ${{\bm \xi}^0}$, and the probability of a particular set of initial conditions, respectively. For deterministic evolution, the conditional probability is a delta function, which maps the initial conditions to the final state at time $t$. Further, the output ${\cal O}$ is a function of ${\bm \xi}(t)$, and will have a distribution that is derived from $P({\bm \xi}^0)$.

Type II events are at the root of some of the common anomalous behavior in combustion systems. High-altitude relight is a relevant example, where the inability to know the exact flow field as well as ignitor conditions leads to variability in ignition times. Another Type II example is soot formation in gas turbine combustors. The initial conditions can have tremendous effect on the history of individual soot particles. Due to the nature of soot formation, only a select set of initial conditions lead to the promotion of soot formation. Consequently, when ${\cal O}$ is directly connected to soot statistics, its PDF could have a non-compact or broad support. More importantly, only select realizations (i.e., initial conditions of the dynamical system) will contribute significantly to the statistics. This requires that the simulation tools are able to sample realizations that lead to these extreme events.

\paral{Type III Events - Attractor-driven Rare Events}

Type III events are related directly to the inability to fully characterize the attractor of a high-dimensional dynamical system. In general, such events are associated with parts of phase-space that are not traversed often, or that are not encountered due to the operational nature of the combustion device. Consequently, such events are rare and occur with low probability. 

In order to describe these events, the phase-space of the dynamical system can be partitioned into these sub-volumes:
\begin{itemize}
    \item Realizable space ${\cal A}_R$ is defined as the volume of state-space that allows thermodynamically consistent set of values for the phase-space variables. In general, ${\cal A}_R$ is a large volume, and its boundaries are unknown due to the high-dimensionality of the state-space. While realizability is guaranteed in a physical experiment, rigorous definition of thermodynamic constraints is necessary for numerical experiments. It is assumed here that all operations involving the dynamical system given above are subject to these unstated but precise thermodynamic constraints. 
    \item Accessed space ${\cal A}_S$ is defined as the region of state-space that is traversed in a given simulation. This volume depends on ${\cal N}$, ${\cal P}$, and ${\bm \xi}^0$, but also depends on the time for which the dynamical system evolves. In other words, the volume is an increasing function of time.
    \item Attractor ${\cal A}$ defines the irreducible volume of the attracting set in which the long-term dynamics of the system are contained. In chaotic systems, these regions form a so-called strange attractor \cite{eckmann_ruelle}.
\end{itemize}

\begin{figure}[ht]
\centering
\includegraphics[width=0.55\textwidth]{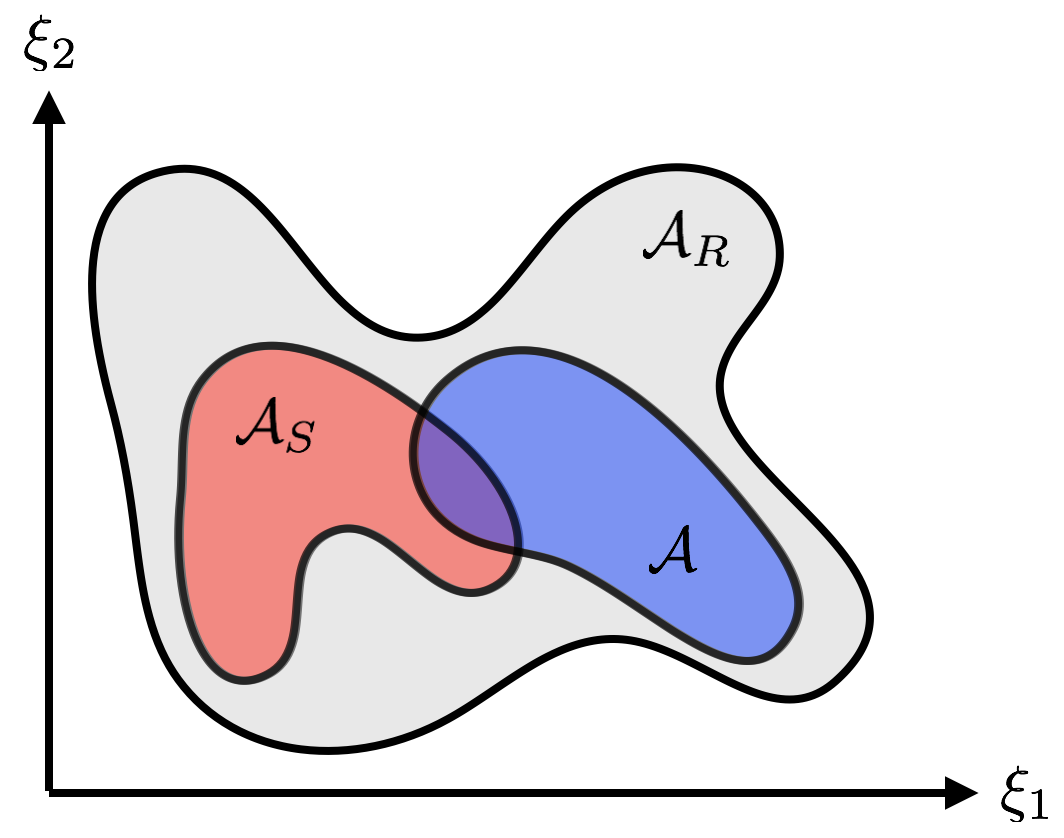}
\caption{Illustration of the phase-space partitioned between the realizable space $\mathcal{A}_R$, the accessed space $\mathcal{A}_S$ and the attractor $\mathcal{A}$.}
\label{fig:space}
\end{figure}

Figure~\ref{fig:space} shows a schematic of the three sub-volumes defined above. The classification of rare events starts with the fact that such events are possible due to the differences in the volumes occupied by ${\cal A}_S$ and ${\cal A}$. In other words, the attractor can be distinct from the accessed space. One common reason is that if the initial conditions are not on the attractor and/or the system is operated only for a finite time (such as the ignition process), the trajectory may not encompass all of the attractor space. Based on these volumes, the following sub-classes can be defined:
\begin{itemize}
\item Type IIIA Events: Black Swans - The term ``Black Swan" \cite{taleb} has been related to the fact that until European explorers reached Australia, only white swans were known to exist. The finding of black swans in Australia invalidated the assumption that all swans are white in color. In the current context, this term refers to events that are pre-existing in the system, but has not been encountered yet. This can arise due to the fact that ${\cal A}_S$ coincides with only a small part of ${\cal A}$ due to finite operation time, as well as limited changed in initial conditions. The anomalous event may be associated with a low-probability region of phase-space that is not accessed unless the system is operational for a very long time, or the system is operated with vastly different initial conditions each time (this is not always possible in physical systems). Although the attractor traps the long-time dynamics of the system, the time taken to travel along the attractor is not known. Consequently, some sections of the attractor may be sparsely traversed, leading to the Black Swan events if these regions also cause anomalous behavior in the output variables ${\cal O}$. An example of such configuration can be obtained using the Lorenz 63 attractor \cite{lorenz} shown in Figure.~\ref{fig:type3a}. Only part of the attractor is traversed if the trajectory is not run for a long enough. Starting from two different initial conditions can then lead to explore two different part of the attractors. 

\begin{figure}[ht]
\centering
\includegraphics[width=0.55\textwidth]{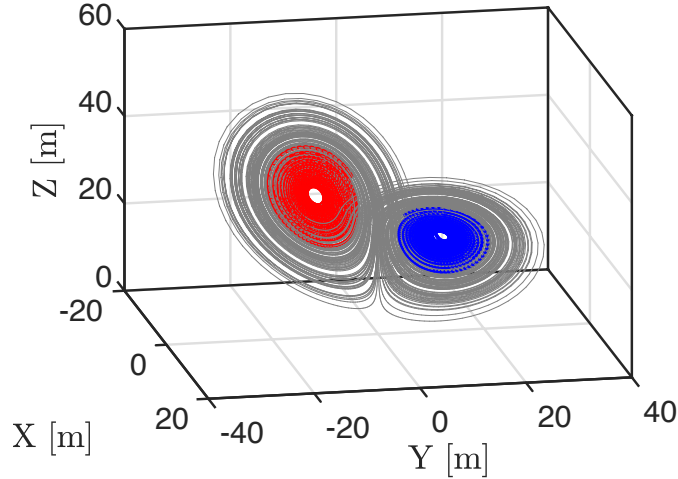}
\caption{Incomplete exploration of the Lorenz 63 attractor. Trajectory started from (x,y,z)=(-5.714,-8.414,26.68) (red thick line); from (5.714,8.414,26.68) (blue thick line). Both ran for 15s with parameters ($\rho$, $\sigma$, $\beta$) = (28,10,8/3). Global attractor (grey line).}
\label{fig:type3a}
\end{figure}

\item Type IIIB Event: Disjoint attractor jumps - The system described has again all its properties (including the attracting set) defined by $\mathcal{N}$ and $\mathcal{P}$. This discussion treats the case where $\mathcal{N}$ and $\mathcal{P}$ and \textit{a fortiori}  the attracting set are held fixed. In the case of non ergodic systems, the attracting set of the system can contain multiple disjoint attractors \cite{ruelle81}. In this case the attracting set for the dynamical system is not a single connected region (connected by some solution orbit) anymore. Then, there exists many different islands of disconnected regions in phase-space, each acting as a basin of attraction for the individual attractors. In such cases, the choice of initial conditions can lead to the system being trapped in different attractor regions, and even for long operational times, the system may not switch between elements of the attracting set. Note that this is different from multistable systems which oscillate between near-constant QoI. In the multistable cases, a single attractor contains the oscillatory motion \cite{tian_bistable}. A Black Swan case that appears to have ${\cal O}$ sensitive to initial conditions ${\bm {\xi}}^0$ can be falsely interpreted as a Type IIIB system since the orbit linking the attractors has not been seen yet.

In this type of system, a rare event can come from a sudden transition from one attractor to the other. This could be achieved by a sudden displacement of the trajectory, for instance, through addition or removal of energy. This might cause the system to move to the basin of attraction of a different attractor. If such event were to happen, the solution would remain in a new phase-space region for long period of time. Again this would be considered a relevant event only if ${\cal O}$ appears to be sensitive to the change of attractor. Such an event is represented in Fig.~\ref{fig:type3b}.

An example of Type IIIB event is the use of a spark to ignite a flame. Without this source of energy, which is provided only temporarily, the system will exist in the non-reactive section of phase-space. The spark creates an orbit pathway in the phase-space that links the non-reactive part to the reactive part of the phase-space for the same ${\cal N}$ and ${\cal P}$. Once the system ignites, without a change in the operating or boundary conditions, it may never reach back the non-reactive portion of phase-space. 

\item Type IIIC Event: Changing attractor - These events are the most pernicious of rare events in that they invalidate the underlying assumption leading the dynamical system description characterized by solely ${\cal N}$ and ${\cal P}$. Such outcomes are tied to perturbations to the system that changes the structure of the attractor. For instance, a change in the shape of the combustor due to excessive heat loads, or the introduction of additional physics that did not exist in the original system. In the former case, ${\cal P}$ is altered even if ${\cal I}$ is held constant, which could change the structure of the attractor. In the latter case, the operator ${\cal N}$ does not contain the entire set of physics, and should be considered as a reduced set of governing ODE. The dimensionality of the real system is larger than the simulated phase-space, allowing the real system to move to parts of phase-space that are forbidden by the assumed structure of ${\cal N}$. Again, such changes can be temporary, time-varying, or permanent, depending on the nature of the perturbation to the system. The main distinction with the Type IIIB event is that the change in $\mathcal{O}$ does not stem from a solution displacement, but from a modification of the stability properties of the system. Annihilating such an event would require to change the attractor and if needed, displace the solution. This is different from a Type IIIB event which would only require to displace the solution back to its original orbit. The exogenous shock, which causes these changes, is of an unquantified nature, which makes the simulation and prediction of such events nearly infeasible. Nevertheless, it is possible to improve robustness of systems by testing the response of the dynamical system to shocks of a known nature.
\end{itemize}

\begin{table}[h]
\centering
\caption{Classification of Anomalous Events.}
\begin{tabularx}{0.99\textwidth}{|*{6}{>{\centering\arraybackslash}X|}}
\hline
 & Type I & Type II & Type IIIA & Type IIIB & Type IIIC \\ \hline
Trigger & $\mathcal{I}$ & $\delta {\bm {\xi}}^0$ & none & trajectory displacement & $\delta \mathcal{P}$ or $\delta \mathcal{N}$ \\ \hline
Attracting set & fixed & fixed & fixed and large & fixed with disjoint attractors  & changes $|\frac{\delta \mathcal{A}}{\delta \mathcal{N}}| > 0$, or $|\frac{\delta \mathcal{A}}{\delta \mathcal{P}}| > 0$, \\ \hline
Output sensitivity & $\mathcal{O}(\bm \xi, \mathcal{I}) \approx \mathcal{O}(\mathcal{I})$ & $|\frac{\delta \mathcal{O}}{\delta {\bm \xi}}| \gg 1$ & $\exists {\bm \xi}_c \in \mathcal{A},$ 

$|\frac{\delta \mathcal{O}}{\delta \mathcal{\bm \xi}}|_{{\bm \xi}_c}| \gg 1$ & $|\frac{\delta \mathcal{O}}{\delta \mathcal{A}_i}| \gg 1$ & $|\frac{\delta \mathcal{O}}{\delta \mathcal{A}}| \gg 1$ \\
\hline
\end{tabularx}
\label{table:Example}
\end{table}

\section{Role of Classification in Developing Simulation Tools}
\label{sec:method}

From the classification of the origin of the anomalous events, some requirements about the computational and the experimental tools can be formulated. The most impactful events arise from Type III processes. Those events introduce previously unseen trajectories or even convergence for the system. Therefore, such events are also the most difficult to experimentally observe or computationally simulate. A useful way forward is to consider these dynamical events as a hierarchy, with the Type I events providing the most information regarding the physics of such anomalous behavior. In this sense, computational tools and theoretical ideas can be developed based on these events, and progressively expanded to capture Type III events. We adopt this methodology here.

\subsection{Type I events}
The ability of predicting Type I events requires to design a system responding perfectly to a change of a set of input parameters $\mathcal{I}$. An experimental set up able to trigger on demand an anomalous event is the right candidate to validate Type I related predicting tools. The current work done in combustion already addresses the modeling of several of such anomalous events. Combustion instabilities such as premixed flame flashback \cite{BLF_chann} or thermoacoustic driven instabilities \cite{poinsot_accoustic} have been successfully predicted in the past. In general, capturing the mean response of a system mainly driven by the input parameter is the goal of existing modeling approaches.

\subsection{Type II events}
Designing predictive tools and validation experiments for Type II events is not as straightforward. The predictive tool has to be able to capture the large sensitivity of some QoI with respect to individual phase-space trajectories (or solution history). It requires to a) observe an outcome sensitive enough to instantons, b) be able to correlate the outcome variations to the instanton. It can create new numerical challenges but also experimental complications if the trajectory variation is small. Promising QoI for the prediction of Type II events in combustion are related to slow processes which are dependent on the solution trajectory (soot formation) or phase-space location sensitive processes (ignition or acoustic instabilities \cite{gotoda}). 

While type II events appear to be closely related to the field of uncertainty quantification, there is an important distinction. In the problem of uncertainty propagation, prior information in propagated through the system. Since the prior makes an assumption about the importance of specific phase-space locations (through a PDF), the outcome is also biased by towards trajectories that originate from highly weighted initial conditions. Since rare events may not be associated with these locations, such a propagation process will only capture the mean outcomes. Here, the focus is on predicting the tails of the outcomes, which will require considerable innovation in algorithms. Recently, such methods have begun to emerge, albeit for small-dimensional systems. They are divided in two main classes of methods: the important sampling method \cite{rubino_tuffin,gudmundsson_thesis} and the importance splitting method \cite{cerou,wouters_bouchet,ffs}. Although promising results can be obtained for small dimensional configurations \cite{AMS_md}, application to discretized continuous fields is still an open research problem. Recently, such techniques have been used for combustion applications \cite{hassanaly_lE}. 

\subsection{Type III events}
The last type of event is the most difficult to predict since it involves non previously observed locations of phase-space. Such events can be captured only through an attracting set search process. For the Type IIIA event, either different initial conditions could be used, or the system could be run for a very long time. Another method could be to numerically change the dynamics of the system and wisely displace the solution inside the attractor knowing its local structure. That way the motion on the attractor could be accelerated. Recently, Tailleur \cite{tailleur} proposed to use importance splitting methods to look for critical points ${\bm \xi}_c$ on the attractor. Finally, designing Type IIIC events would require the analysis of transient changes due to a modification of operating conditions. This introduces the concept of the sensitivity of an attractor which was originally considered as an invariant set \cite{ruelle81}. Some recent work \cite{qiqi_lss} considered the dependency of the attractors with $\mathcal{P}$ but additional efforts are required for the analysis of the solution transition during these changes.

\section{Conclusion}

In an attempt to lay the foundations of anomalous event predictions in turbulent combustion, a classification is established in relation to the governing equations. The root of anomalous events are conceptually described and underline the fact that distinct properties need to be captured for each one of the anomalous event types. It is in particular noticed that anomalous behaviors in the neighborhood or far from known phase-space trajectories can be explained by different mechanisms. It is hoped that the understanding of this mechanism using a deterministic description of the anomalous event will lead to advanced methodologies for the search and control of rare events, thereby fulfilling an important goal of predictive simulations.



\bibliographystyle{mcs10}
\bibliography{bibtex_database}

\end{document}